\documentstyle[12pt]{article}
\begin{document}
\setcounter{page}{0}
\thispagestyle{empty}

\begin{center}

\vfill

{\Large On calculation of the off-shell renormalization functions in
the $R^2$-- gravity.}
\vspace{1cm}

{\large M.~Yu.~Kalmykov}
\footnote {E-mail: $kalmykov@thsun1.jinr.dubna.su$}.
and
{\large D.~I.~Kazakov}
\footnote{E-mail:  $kazakovd@thsun1.jinr.dubna.su$},
\vspace{1cm}

{\em Bogoliubov Laboratory of Theoretical Physics,
     Joint Institute for Nuclear  Research,
     $141~980$ Dubna $($Moscow Region$)$, Russian Federation}

\end{center}
\vspace{1cm}

\begin{abstract}
A new way how to calculate the off-shell renormalization functions
within the $R^2$-gravity has been proposed. The one-loop
renormalization group equations in the approach suggested have been
constructed. The behaviour of effective potential for an massless
scalar field interacting with the quantum gravitational field has been
analyzed in this approach.
\end{abstract}

\pagebreak

As known, the Einstein theory of gravity is non-renormalizable
\cite{Einstein}. Whe\-re\-as the theory with terms which are quadratic
in the curvature tensor is renormalizable in all orders of the perturbation
theory \cite{ren} but not unitary \cite{BOS} (ghosts and tachyons are present
in its spectrum ).  So, this $R^2$-gravity
cannot be accepted as a basic theory.  Nevertheless, this theory may
be considered as a model for studying the quantum gravitational
effects. In particular, in the framework of this theory we can use
the renormalization group method.

One of the problems in the $R^2$-gravity is related to the
$\beta_G$-function calculation ($G$ is the Newtonian constant).
$\beta_G$ calculated by a standard way is dependent on the gauge and
parametrization \cite{trace}.  It is commonly believed that this
dependence can be explained if $G$ is thought of as an inessential
coupling constant which does not enter the renormalized $S$-matrix
definition. However, we think of such a dependence as a flaw in the
calculation method.  One should remember that $G$ is a quantity which
can be measured in experiments and enters the classical gravitational
potential definition \cite{Stelle}.  The loop corrections to the
gravitational potential are proportional to the $\beta_G$-function.
It leads to the gauge and parametrization dependence of physical
quantities. Hence, in our opinion, the $\beta_G$-function  should not
depend on the gauge and parametrization. For getting the
$\beta_G$-function which is independent on the gauge and
parametrization we should base on some additional suggestions in the
framework of the traditional procedure or should calculate new
objects like the Vilkovisky-DeWitt effective action \cite{Avramidi}.
However, in the Vilkovisky-DeWitt formalism there is an ambiguity in
the choice of the configurational space metric for the quantum
gravity. For this reason $\beta_G$ will depend on this metric.  In
this paper we suggest a new way for this problem solution.  The
method suggested for finding the correct $\beta_G$-function is based
on putting the non-zero renormalization constant on the metric field
\footnote{The non-zero renormalization constant for the metric field
in $(2+\varepsilon)$-gravity has been considered in refs \cite{2D}.}.

Let us consider the $R^2$-gravity with Lagrangian
\footnote{We use the following notations:
$$ c = \hbar = 1;~~~~~ \mu , \nu  = 0,1,2,3;~~~~~
(g) = det(g_{\mu \nu }), ~~~~~\varepsilon  = \frac{4-d}{2}
$$
$$
R^\sigma _{~\lambda  \mu  \nu } = \partial_\mu \Gamma^\sigma
_{~\lambda \nu }  - \partial_\nu \Gamma^\sigma _{~\lambda \mu }
+ \Gamma^\sigma_{~\alpha \mu } \Gamma^\alpha_{~\lambda  \nu }
- \Gamma^\sigma_{~\alpha  \nu }  \Gamma^\alpha_{~\lambda \mu },~~~~~
R_{\mu \nu } = R^\sigma_{~\mu \sigma \nu },~~~~~
R =  R_{~\mu \nu } g^{\mu \nu } $$
where $\Gamma^\sigma_{~\mu \nu } $ is the Riemann connection and
$W^2 = R_{\mu \nu}^2 - \frac{1}{3}R^2, \nabla^2 = g^{\mu \nu}
\nabla_\mu \nabla_\nu$}

\begin{equation}
L_{GR} = \Big(
\frac{1}{\lambda} W^2 - \frac{\omega}{3\lambda} R^2
- \frac{R}{k^2} + \frac{2 \sigma}{k^4} \Big) \sqrt{-g}
\label{theory}
\end{equation}

\noindent
where $\lambda, \omega$ and $\sigma$ are the dimensionless constants,
$k^2 = 16 \pi G$.
The space-time we work is topologically trivial, without a boundary,
the Euler number equals zero. Thus, we have a right to use the
relation:  $R_{\mu \nu \sigma \lambda}^2 = 4 R_{\mu \nu}^2 - R^2$.

The theory described by the action (\ref{theory}) is multiplicatively
renormalizable in all orders of the perturbation theory.  The
calculation of Green functions as well as $S$-matrix elements
including radiative corrections can be carried out in the framework
of the theory under consideration. For getting the finite Green functions
in the standard field theory not only physical parameters but fields
themselves also should be renormalized. As to the $S$-matrix elements
the result has to be the same as when fields are not renormalized.
Following the arguments of \cite{Abbott} it can be shown that the
renormalization of quantum and ghost fields in the background field
method is not essential to the one-loop background Green
functions calculation. Consequently, at the one-loop level, in the
theory (\ref{theory}) the five renormalization constants may arise:
for the physical parameters $Z_\lambda, Z_\omega, Z_G, Z_\sigma$ and
for the background metric field $Z_g$.  All one-loop singularities of
the theory can be absorbed into these constants. In the MS-scheme
all $Z_i$ constants contain only poles in $\varepsilon$.

Let us consider how the one-loop renormalization group equations will
be modified due to the non-zero renormalization constant for the
metric field.  Let the tensor density $g_{\mu \nu}^\ast = g_{\mu \nu}
(-g)^r$ (where $r \neq - 1/4$) play the role of a dynamical variable.
The one-loop counterterms in the background field method have the
form:
\footnote{For the sake of completeness we present the results $\{
\theta_i \}$ in the trivial parametrization $r=0$ and in the minimal
gauge given in the paper \cite{Avramidi}
$$\theta_2 =  \frac{1}{16 \pi^2} \frac{133}{20}, ~~~~~~
\theta_3 = \frac{1}{16 \pi^2} \Big( \frac{5}{3}\omega^2
+ \frac{5}{2} \omega + \frac{5}{24} \Big), ~~~~~
\theta_4 = \frac{1}{16 \pi^2} \Big( \frac{5}{3}\omega
- \frac{13}{12}  - \frac{1}{8 \omega} \Big) \lambda,
$$
$$
\theta_5 = \frac{1}{16 \pi^2} \Big( \frac{5}{4} \lambda^2
+ \frac{\lambda^2}{16 \omega^2}  + \frac{28}{3} \sigma \lambda
+ \frac{1}{3} \frac{\sigma \lambda}{\omega} \Big)
$$
}

\begin{equation}
\Gamma_{div}^{GR} = \frac{1}{\varepsilon}
\int
\left(\theta_2 W^2 + \frac{\theta_3}{3} R^2
+ \theta_4 \frac{R}{k^2} + \frac{\theta_5}{k^4} \right) \sqrt{-g}
~d^4 x.
\label{1loop}
\end{equation}

Supposing that after multiplicative redefinition of the metric
and parameters

\begin{eqnarray}
g_{\mu \nu}^* & \to & g_{\mu \nu}^{*B} =  Z_g g_{\mu \nu}^{*R} =
\left(1 + \delta Z_g \right) g_{\mu \nu}^{*R},
\nonumber \\
\lambda & \to & \lambda_B = Z_\lambda \lambda_R =
\left( \lambda_R + \delta Z_\lambda \right),
\nonumber \\
\omega & \to & \omega_B = Z_\omega \omega_R =
\left( \omega_R + \delta Z_\omega \right),
\nonumber \\
\sigma & \to & \sigma_B = Z_\sigma \sigma_R =
\left( \sigma_R + \delta Z_\sigma \right),
\nonumber \\
G & \to & G_B = Z_G G_R =
\left( 1 + \delta Z_G \right) G_R
\end{eqnarray}

\noindent
the one-loop background Green functions obtained from the effective
action $\Gamma_R = \Gamma_B^{GR} - \Gamma_{div}^{GR}$ should be
finite at $\varepsilon \to 0$ we have the following system of
equations for $\delta Z_i$

\begin{eqnarray}
\delta Z_\lambda & = & - \frac{\theta_2}{\varepsilon} \lambda^2_R,
\label{Zl}
\\*
\delta Z_\omega & = & - \frac{\theta_3 + \theta_2 \omega_R}{\varepsilon}
\lambda_R,
\label{Zo}
\\*
\delta Z_\sigma  & = & \frac{1}{2}
\frac{\theta_5 + 4 \theta_4 \sigma_R }{\varepsilon},
\label{Zs}
\\*
\delta Z_G  - \frac{1}{s} \delta Z_g & = & \frac{\theta_4}{\varepsilon}
\label{Z}
\end{eqnarray}

\noindent
where $s \equiv 4r+1 \neq 0$ and we take into account that

$$
g^{\mu \nu} \to \left(1 - \frac{1}{s} \delta Z_g \right) g^{\mu \nu},
~~~~~
\sqrt{g} \to \left(1 + \frac{2}{s} \delta Z_g \right) \sqrt{g}.
$$

The renormalization constant for the metric field appears only in the
terms $R \sqrt{-g}/k^2 $ and $\sqrt{-g}/k^4$. It is conditioned by
the fact that the tensor $R^\sigma_{~\mu \lambda \nu}$ is built out
of the combinations $\phi^{-1} \partial \phi$ and is invariant under
multiplicative redefinition of the fields.  Moreover, at the same time the
combination $g^{\mu \nu} g^{\alpha \beta} \sqrt{-g}$ is invariant
under multiplicative redefinition of the metric. So, in the gravity
compared with the standard field theory the renormalization constant
for the field is defined by low powers of the kinetic term.

In order to find a definite solution for the equation (\ref{Z}) we
need some additional conditions. We don't know exactly how to find
new equation. We suggest that such an additional equation can be
obtained in the on-shell approach.  We suppose that all on-shell
divergences can be removed by the redefinition of only physical
parameters (the gravitational and dimensionless constants).  This
assumption is based on two points:

\begin{itemize}
\item
S-matrix in the background field method is identical to the
conventional S-matrix \cite{S-matrix}
\item
for the renormalized S-matrix elements calculation we should
renormalize only physical parameters.
\end{itemize}

\noindent
As known, for the $R^2$--gravity we may limit ourselves to using only the
trace of the motion equation \cite{trace2}

\begin{equation}
g^{\ast \mu \nu}_B \frac{\delta L_{GR}}{\delta g^{\ast \mu \nu}_B}
 = \left( - \frac{s}{k^2_B} \Big( R_B - \frac{4 \sigma_B}{k^2_B}
\Big) + 2 s \frac{\omega_B}{\lambda_B} \nabla^2 R_B \right)
\sqrt{-g_B} \equiv 0.
\label{equation}
\end{equation}

Taking into account (\ref{Zl}--\ref{equation}), we have

\begin{eqnarray}
\delta Z_G & = & 0,
\label{ZG}
\\
\delta Z_g & = & -s \frac{\theta_4}{\varepsilon}.
\label{Zh}
\end{eqnarray}

\noindent
It is easy to show that in our approach $\delta Z_G = 0$ in all
orders of the perturbation theory. The one-loop counterterms are not
polynoms in Lagrangian parameters (see the footnote 5 at the page 2).
It means the smallness of $\delta Z_i$ is provided by the expansion
over constants $\hbar$ (loop expansion). In this case in order to
calculate generalized $\beta$-functions we can use the equations
>from \cite{Kazakov}. Introducing the definition $\mu^2 \frac{d}{d
\mu^2} g_{Bare} = - \varepsilon g + \beta_g $, we obtain that in the
MS-scheme at the one-loop level

\begin{equation}
\beta_i = \delta Z_i \varepsilon.
\label{beta}
\end{equation}

\noindent
At the one-loop level the coefficients $\theta_2, \theta_3$ and
the combination $\theta_5 + 4 \sigma_R \theta_4$ are independent on the gauge
and parametrization off-shell \cite{independence}. Using the
relations (\ref{Zl}) - (\ref{Zs}), (\ref{ZG}) - (\ref{beta}), we get that
the one-loop $\beta$--functions for physical parameters are also independent on
the gauge and parametrization. $\beta_G=0$ in all orders of the perturbation
theory. So, in the space without boundaries and interactions with the matter
fields there are no loop corrections to the Newtonian constant. As to the
coefficient $\theta_4$ and the one-loop anomalous dimension of the
metric field $\gamma_g$ they depend on the gauge and parametrization
\cite{trace}.  It does not contradict to the basics of the quantum
theory. The introduction of the non-zero renormalization constant for the
metric field allows also to explain the dependence of one-loop results from
the configurational space metric in the Vilkovisky-DeWitt formalism.
The standard form for the configurational space metric in the gravity is

$$
\gamma^{\mu \nu \alpha \beta} = \Big( g^{\mu \alpha} g^{\nu \beta}
+ g^{\mu \beta} g^{\nu \alpha} - {\bf a}
g^{\mu \nu} g^{\alpha \beta} \Big) \sqrt{-g}
$$

\noindent
In order to fix {\bf a} it is required \cite{EA} that the metric
should be fixed in the space of fields in accordance with the
classical action coefficients at the highest space-time derivatives.
The calculations performed for the quantum gravity in the
Vilkovisky-DeWitt formalism lead to a dependence of
physical quantities from {\bf a} \cite{metric}.  At present this
question (correct choice of the configurational space metric in the
gravity) is still open.  By using the method for the renormalization
functions calculation suggested in this work within the Vilkovisky-DeWitt
formalism, we have $\beta_G = 0$ and all the dependence from {\bf a} is
absorbed into the anomalous dimension of the metric field $\gamma_g$.
This dependence of $\gamma_g$ from {\bf a} can be explained in
precisely the same way as the gauge and parametrization dependence of
the field anomalous dimension in the standard quantum field theory.

It can be manifested that the multiplicative renormalization of the metric
field is related to only the conformal mode. The easiest way to do it
is to choose a conformal parametrization where dynamical variables
are $\psi_{\mu \nu} = g_{\mu \nu} (-g)^{-\frac{1}{4}}$ and $\pi =
(-g)^{\frac{m}{4}}$, $m \neq 0,~~ det \psi_{\mu \nu} = 1$.
In this parametrization the fields $\psi_{\mu \nu}$ and
$\pi$ are considered to be independent dynamical variables. As a
consequence, in the general case there are two different
renormalization constants $Z_\psi, Z_\pi$ for the fields $\psi$
and $\pi$ respectively. The similar arguments we have used above
result in the following system for the one-loop renormalization
constants $\delta Z_i$ definition

\begin{eqnarray}
-\frac{\delta Z_\lambda}{\lambda^2} - 2 \delta Z_\psi & = &
\frac{\theta_2}{\varepsilon},
\nonumber \\*
\frac{\delta Z_\omega}{\lambda}
- \frac{\omega}{\lambda^2}\delta Z_\lambda
- 2 \frac{\omega}{\lambda} \delta Z_\psi
& = & \frac{\theta_3}{\varepsilon},
\nonumber \\*
\delta Z_\sigma  - 2 \left( \delta Z_G - \frac{1}{m} \delta Z_\pi
\right)\sigma_R & = & \frac{1}{2} \frac{\theta_5}{\varepsilon},
\nonumber \\*
\delta Z_G  + \delta Z_\psi - \frac{1}{m} \delta Z_\pi & = &
\frac{\theta_4}{\varepsilon}
\nonumber
\end{eqnarray}

\noindent
where we use the relations
$\psi^B_{\mu \nu}  = \left(1 + \delta Z_\psi \right) \psi_{\mu
\nu}^R$ and
$\pi_B  = \left(1 + \delta Z_\pi \right) \pi_R$.
The conditions of renormalizability on-shell give some additional
equations for the renormalization constants definition.  As a result,
the old solutions (\ref{Zl})--(\ref{Zs}), (\ref{ZG}) are obtained
once more, the solution (\ref{Zh}) will be replaced with

\begin{eqnarray}
\delta Z_\psi & = & 0,
\label{ZP}
\\
\delta Z_\pi & = & -m \frac{\theta_4}{\varepsilon}.
\label{Zp}
\end{eqnarray}

Let us consider the $R^2$--gravity interacting with an massless scalar
field. The gravitational field Lagrangian has the form
(\ref{theory}), the Lagrangian for the scalar field is

\begin{equation}
L_{mat} = \left(
\frac{1}{2} g^{\mu \nu} \partial_\mu \varphi \partial_\nu \varphi
- \frac{\rho \varphi^4}{4 !} + \frac{1}{2} \xi R \varphi^2 \right)
\sqrt{-g}
\label{matter}
\end{equation}

\noindent
where $\rho, \xi$ are the dimensionless constants. Let the tensor
density $g_{\mu \nu}^\ast = g_{\mu \nu}(-g)^r$, where $r \neq - 1/4$
and scalar density $\phi = \varphi (-g)^\chi, \chi$ is an arbitrary
number be dynamical quantities.  The one-loop divergencies in the
background field method are $\Gamma_{div} = \Gamma_{div}^{GR} +
\Gamma_{div}^{mat}$, where the functional form of $\Gamma_{div}^{GR}$
is given in (\ref{1loop}), and

\begin{equation}
\Gamma_{div}^{mat} = \frac{1}{\varepsilon} \int
\left(
\gamma_1 \frac{1}{2} g^{\mu \nu} \partial_\mu \varphi \partial_\nu
\varphi - \gamma_2 \frac{\varphi^4}{4 !}
+ \gamma_3 \frac{1}{2} R \varphi^2
\right) \sqrt{-g}~d^4 x.
\label{2loop}
\end{equation}

\noindent
Supposing that all divergencies can be removed by the renormalization of
constants and fields, the following equations system for the renormalization
constants is obtained in addition to
(\ref{Zl})--(\ref{Zs}), (\ref{ZG}), (\ref{Zh})
\footnote{For this theory one needs to use the following equation
$$ g^{\ast \mu \nu}_B \frac{\delta L}{\delta g^{\ast \mu
\nu}_B} = \left( - \frac{s}{k^2_B} \Big( R_B - \frac{4
\sigma_B}{k^2_B} \Big) + 2 s \nabla^2 \left(
\frac{\omega_B}{\lambda_B} R_B + \frac{6 \xi -1}{8} \varphi^2_B
\right) \right) \sqrt{-g_B} \equiv 0.  $$ As a consequence, the eqs.
(\ref{ZG}) and (\ref{Zh}) do not change their functional form.}

\begin{eqnarray}
\delta Z_\xi & = & \frac{\gamma_1 \xi - \gamma_3 }{\varepsilon},
\label{Zx}
\\*
\delta Z_\rho & = & \frac{2 \gamma_1 \rho - \gamma_2}{\varepsilon},
\label{Zr}
\\*
\delta Z_\phi  & = & - \frac{1}{2} \frac{\gamma_1}{\varepsilon}
- \frac{1-8\chi}{2s} \delta Z_g.
\label{Zphi}
\end{eqnarray}

\noindent
where
$\phi_B  = \left(1 + \delta Z_\phi \right) \phi_R$,
$\xi_B  = \left(\xi_R + \delta Z_\xi \right)$ and
$\rho_B  = \left(\rho_R + \delta Z_\rho \right)$.
In the conformal parametrization the relations (\ref{Zr}),
(\ref{Zphi}) should be replaced with the following

\begin{eqnarray}
\delta Z_\rho & = & \frac{2 \gamma_1 \rho - \gamma_2}{\varepsilon}
- 2 \rho \delta Z_\psi,
\nonumber
\\*
\delta Z_\phi  & = & - \frac{1}{2} \frac{\gamma_1}{\varepsilon}
- \frac{1-8\chi}{2m} \delta Z_\pi + \frac{1}{2} \delta Z_\psi
\nonumber
\end{eqnarray}

\noindent
where $\delta Z_\psi$ and $\delta Z_\pi$ are defined in (\ref{ZP})
and (\ref{Zp}) respectively. The renormalization constant for the
scalar field is related to the corresponding constant for the metric
field if $\chi\neq 1/8$. So, in our approach the renormalization
group equation for the scalar field constant change its form, and
the non-zero renormalization constant for the metric field appears.

Let us analyze the renormalization group for the improved effective
potential \cite{potential}.  The renormalization group equation for
the effective potential in our approach has the following form
\begin{equation}
\left( \mu^2 \frac{\partial}{\partial \mu^2} + \beta_\rho
\frac{\partial}{\partial \rho} + \beta_\xi \frac{\partial}{\partial
\xi} - \gamma_\phi \phi \frac{\delta}{\delta \phi} + \gamma_g g^{\ast
\mu \nu} \frac{\delta}{\delta g^{\ast \mu \nu}} \right) V = 0
\label{rg}
\end{equation}

\noindent
where $\beta_\rho, \beta_\xi, \gamma_\phi $ and $\gamma_g$ are the
renormalization group functions; $\phi$ is the scalar density and
$g^{\ast \mu \nu}$  is the tensor density respectively. We consider
$R_{\mu \nu}$ as a kinetic term. As a consequence, the derivative
$\frac{\delta}{\delta g^{\ast \mu \nu}}$ does not influence on $R_{\mu
\nu}$, but only on $g^{\mu \nu} \sqrt{-g}$ and $\sqrt{-g}$. Splitting
the potential $V$ into two parts ($V_1$ which is independent on the curvature
and $V_2$ linear in the curvature), we obtain two equations

\begin{equation}
\left(
\mu^2 \frac{\partial}{\partial \mu^2}
+ \beta_\rho  \frac{\partial}{\partial \rho}
- 4 \gamma_\phi
- 2 \frac{1-8\chi}{s} \gamma_g \right) V_1 = 0
\label{A}
\end{equation}

\begin{equation}
\left(
\mu^2 \frac{\partial}{\partial \mu^2}
+ \beta_\xi \frac{\partial}{\partial \xi}
- 2 \gamma_\phi
- \frac{1-8\chi}{s} \gamma_g \right) V_2 = 0
\label{B}
\end{equation}

\noindent
The solutions of eqs. (\ref{A}) and (\ref{B}) are the following:

\begin{equation}
V (t) =
\frac{\rho(t) \varphi^4}{4 !} \sqrt{-g} f^4(t)
- \frac{1}{2} \xi (t) R \varphi^2 \sqrt{-g} f^2(t)
\label{solution}
\end{equation}

\noindent
where

\begin{eqnarray}
t & = & \ln \frac{\varphi^2 (-g)^{2 \chi_1}}{\mu^2}
\label{t}
\\*
f(t) & = &
\exp \left[ - \int_0^t
\frac{\gamma_\phi(\tau) + \frac{1-8\chi}{2s} \gamma_g(\tau)}
{1+2 \gamma_\phi(\tau) + 8 \frac{\chi_1-\chi}{s} \gamma_g(\tau)}
d \tau \right]
\label{f}
\\*
\frac{d  \rho (t)}{d t} & = &
\frac{\beta_\rho(t)}
{1+2 \gamma_\phi(t) + 8 \frac{\chi_1-\chi}{s} \gamma_g(t) },
~~~~\rho(0) = \rho
\label{rho}
\\*
\frac{d \xi (t) }{d t} & = &
\frac{\beta_\xi(t)}
{1+2 \gamma_\phi(t) + 8 \frac{\chi_1-\chi}{s} \gamma_g(t) },
~~~~\xi(0) = \xi
\label{xi}
\end{eqnarray}

\noindent
and we use the following initial condition

\begin{equation}
V(t=0) = V_{cl} \equiv \left(
\frac{\rho \varphi^4}{4 !} - \frac{1}{2} \xi R \varphi^2 \right)
\sqrt{-g}
\label{inition}
\end{equation}

If $\chi_1 \neq 1/8$ then the solution (\ref{solution}) differs from
a standard one \cite{potential}. However, in the one-loop approximation
where $ f(t) = 1 - \left( \gamma_\phi + \frac{1-8\chi}{2s} \gamma_g
\right) t$ and $\rho(t) = \rho + \beta_\rho t, \xi(t) = \xi +
\beta_\xi t$, the effective potential (\ref{solution}) is the same as the
standard one at arbitrary $\chi$ and $\chi_1$.
It can be explained by the fact that for the field combinations
$\varphi^4 \sqrt{-g}$ and $\varphi^2 g^{\mu \nu} \sqrt{-g}$
entering the effective potential the one-loop renormalization equation in our
approach (see eq. (\ref{Zphi})) has the same form as the traditional one
\cite{potential}.

In the conformal parametrization the renormalization group equation
for the effective potential is

\begin{equation}
\left(
\mu^2 \frac{\partial}{\partial \mu^2}
+ \beta_\rho  \frac{\partial}{\partial \rho}
+ \beta_\xi \frac{\partial}{\partial \xi}
- \gamma_\phi \phi \frac{\delta}{\delta \phi}
- \gamma_\pi \pi \frac{\delta}{\delta \pi}
+ \gamma_\psi \psi^{\mu \nu} \frac{\delta}{\delta \psi^{\mu \nu}}
\right) V = 0
\label{rg1}
\end{equation}

\noindent
The functional form for this solution is the same as for
(\ref{solution}),
where we use the condition $\gamma_\psi = 0$ (see eq.(\ref{ZP})).
The definition of $t$ and the initial conditions are given
in (\ref{t}) and (\ref{inition}). And now

\begin{eqnarray}
f(t) & = & \exp \left[ - \int_0^t
\frac{ \gamma_\phi(\tau) + \frac{1-8\chi}{2 m} \gamma_\pi(\tau)}
{1+2 \gamma_\phi(\tau) + 8 \frac{\chi_1-\chi}{m} \gamma_\pi(\tau)}
d \tau \right]
\\*
\frac{d \rho(t)}{d t} & = &
\frac{\beta_\rho(t)}
{1+2 \gamma_\phi(t) + 8 \frac{\chi_1-\chi}{m} \gamma_\pi(t)},
~~~~\rho(0) = \rho
\\*
\frac{d \xi (t)}{d t} & = &
\frac{\beta_\xi(t)}
{1+2 \gamma_\phi(t) + 8 \frac{\chi_1-\chi}{m} \gamma_\pi(t)},
~~~~\xi(0) = \xi
\end{eqnarray}

This paper shows that in the quantum $4D$ $R^2$--gravity the non-zero
renormalization constant for the metric field may exist and depends
on the choice of gauge and parametrization. Only the conformal metric
mode should be renormalized. In the gravity compared with the
standard field theory the renormalization constant for the field is
defined by low powers of the kinetic term.
The $\beta$--function for physical parameters
obtained in this approach is independent on the gauge and
parametrization at the one-loop level in the MS-scheme in the
dimensional regularization. The $\beta$-function for the Newtonian
constant equals zero in the space-time without a boundary in
all order of the perturbation theory both without and with massless fields
interactions. The renormalization constants for
matter fields acquire a dependence on the metric field renormalization.
In spite of changes in the renormalization group equations the
effective potential for an arbitrary massless theory at the one-loop
level in the linear curvature approximation remains invariable and
is the same as in \cite{potential}.

\noindent
{\bf Acknowledgments}
The authors are grateful very much to L.~Avdeev for fruitful
discussions and useful remarks.  M.~Yu.~Kalmykov is indebted to
S.~Odintsov for discussions of the book \cite{BOS}.

\end{document}